*2024-2025 CRA Quadrennial Paper*

# Prioritizing Computing Research to Empower and Protect Vulnerable Populations

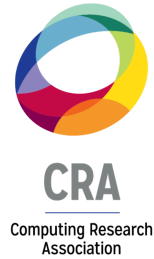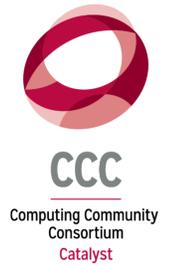

Pamela Wisniewski (Vanderbilt University), Katie Siek (Indiana University Bloomington), Kevin Butler (University of Florida), Gabrielle Allen (University of Wyoming), Weisong Shi (University of Delaware), Manish Parashar (University of Utah)

**Technology can pose significant risks to a wide array of vulnerable populations. However, by addressing the challenges and opportunities in technology design, research, and deployment, we can create systems that benefit everyone, fostering a society where even the most vulnerable are empowered and supported.**

## Technology's Role in Supporting Vulnerable Communities

Vulnerable populations, such as children, older adults, rural populations, people with disabilities, and the LGBTQ+ community, often have intersectional identities that amplify challenges, including disability (affecting 1 in 4 people), domestic violence (1 in 4), and poverty (1 in 9). These issues demand innovative solutions to protect and empower vulnerable individuals. The United States has responded with federal assistance programs to support those in need and demographic-specific policies that mitigate hardships for special groups while benefitting others. For instance, curb cuts in sidewalks (Title II of the Americans with Disabilities Act) benefit people with mobility issues as well as those pushing a stroller or transporting items. Technology accessibility requirements (Section 508 of the Rehabilitation Act of 1973) help people as their abilities change over time or in different contexts (e.g., closed captioning in loud environments). These programs highlight our nation's commitment to supporting vulnerable individuals while benefiting all.

Technical solutions are often seen as low-cost, accessible mechanisms to address vulnerabilities. Technology enables vulnerable populations to acquire knowledge and skills, build communities on social networking platforms, and augment their abilities with assistive technologies. Unfortunately, without proper oversight and development, these same technologies can also pose significant risks. They can perpetuate harm, exploitation, and discrimination, exacerbating existing inequalities and negatively impacting health, social, and economic outcomes for vulnerable individuals and groups. This paper underscores the urgent need to develop ethical, inclusive, and sustainable technologies that empower and protect vulnerable communities. By addressing the challenges and opportunities in technology design, research, and deployment, we can create systems that benefit everyone, fostering a society where even the most vulnerable are empowered and supported.

## Requiring Ethical Practices for Working with Vulnerable Communities

The harms facilitated via technology can scale quickly. For instance, healthcare systems may inaccurately diagnose or skew data for women and people from other underserved groups; scammers may automate financial fraud targeting the elderly and other vulnerable populations; and predictive policing tools may perpetuate social and racial profiling. To address these issues, policies must ensure:

- **Communities have mechanisms to provide feedback** based on their interactions with technologies and researchers.

- **External parties can measure impact** on communities and report back to federal agencies and technology innovators about opportunities and potential impacts/harms. These groups can also help evaluate and mitigate power hierarchies, consider cultural sensitivities, and identify how to maximize benefits while reducing risks.

- **Accountability between communities, innovators, and federal agencies** where the innovators acknowledge feedback and clearly communicate their plans to address findings.

## Systematizing Technology Design Practices for Vulnerable Populations

The level of engagement described above requires substantial effort and investment. Researchers and innovators need incentives (e.g., resources and funding) to engage adequately with vulnerable populations, such as:

- **Open shared infrastructure:** Resources, tools, data, expertise, and best practices must be accessible to all stakeholders, including members of vulnerable communities, ensuring openness and transparency.

- **Community-based policy and governance structure:** Guidance and guardrails must be established for research translation and technology development, considering the longitudinal impacts of technology and retroactively addressing undesired outcomes discovered later.

## Deploying Novel Technologies That Protect and Empower Vulnerable Users

Systematizing design practices should begin early in the training process for innovators and researchers, rather than being addressed only when expanding into specific communities. A two-pronged approach is needed:

- **Enhance training for professionals**, from students to **practitioners**, on working with vulnerable communities ethically and responsibly. Research should focus on addressing human problems, not just technological innovation.

- **Increase opportunities for vulnerable populations** to contribute to the ecosystem, akin to community-based participatory research partnership training programs.

Diverse participation in research improves innovation and results in more inclusive technologies. Reward mechanisms are needed to encourage and support researchers engaged in translational research and technology development. Sustainable funding opportunities and mechanisms should align with the unique requirements of involving vulnerable users. A national community clearinghouse could support the sociotechnical ecosystem, serve as a focal point for engaging with vulnerable populations, and maintain connections with federal agencies foundations, incubators, and social entrepreneurs.

## Taking Accountability to Actively Assess and Mitigate Harm

Policies must ensure accountability for the downstream effects of technological innovations. The unintended consequences of data-driven, artificially intelligent, and surveillance-based technologies — without human oversight — cannot be understated. For example, deep learning-based perception techniques adopted in autonomous vehicles could lead to unexpected decisions that harm users of the transportation system. Similarly, location-tracking apps could imprison individuals in cycles of human trafficking or intimate partner violence.

There is a tension between preserving individual privacy and ensuring national security. This tension must be acknowledged when designing systems that advance national well-being while safeguarding vulnerable populations. New governance models, evaluation frameworks, and policies are needed to ensure that computing research translates into real-world sociotechnical systems that produce long-term net benefits and include safeguards against harm. While innovation should not be hindered, the great power and responsibility of computing research in shaping the health, wealth, and prosperity of society must be recognized.

# Recommendations

- **Develop Ethical and Inclusive Technologies** by prioritizing a national computing research agenda focused on serving and protecting vulnerable communities.

- **Provide Resources, Incentives, and a Shared Infrastructure** to support researchers and innovators engaging with vulnerable populations. Establish a community-based policy and governance structures to guide research and technology development.

- **Enhance Training and Participation** by teaching professionals how to work with vulnerable communities and increasing opportunities for these populations to contribute.

- **Ensure Accountability and Mitigate Harm** by developing new governance models, evaluation frameworks, and policies to ensure that technological innovations produce long-term benefits with safeguards against harm.

---

*This quadrennial paper is part of a series compiled every four years by the **Computing Research Association (CRA)** and members of the computing research community to inform policymakers, community members, and the public about key research opportunities in areas of national priority. The selected topics reflect mutual interests across various subdisciplines within the computing research field. These papers explore potential research directions, challenges, and recommendations. The opinions expressed are those of the authors and CRA and do not represent the views of the organizations with which they are affiliated.*

*This paper was supported by the Computing Community Consortium through the National Science Foundation under Grant No. 2300842 Any opinions, findings, and conclusions or recommendations expressed in this material are those of the authors and do not necessarily reflect the views of NSF.*